\documentclass[fleqn,10pt]{wlscirep}
\usepackage[utf8]{inputenc}
\usepackage[T1]{fontenc}

\newcommand{\onlinecite}[1]{\hspace{-1 ex} \nocite{#1}\citenum{#1}}

\title{Acceleration of the precession frequency for optically-oriented electron 
spins in ferromagnetic/semiconductor hybrids}

\author[1,*]{F. C. D. Moraes}
\author[1,+]{S. Ullah}
\author[2]{M. G. A. Balanta}
\author[2]{F. Iikawa}
\author[3]{Y. A. Danilov}
\author[3]{M. V. Dorokhin}
\author[3]{O. V. Vikhrova}
\author[3]{B. N. Zvonkov}
\author[1]{F. G. G. Hernandez}

\affil[1]{Instituto de F\'{i}sica, Universidade de S\~{a}o Paulo, S\~ao 
Paulo, SP 05508-090, Brazil}
\affil[2]{Instituto de F\'{i}sica ``Gleb Wataghin'', Universidade Estadual de 
Campinas, Campinas, SP 13083-859, Brazil}
\affil[3]{Physico-Technical Research Institute, Nizhny Novgorod State 
University, Nizhni Novgorod 603950, Russia}

\affil[*]{fmoraes@if.usp.br}

\affil[+]{Present Address: Department of Physics, Gomal University, Dera Ismail Khan 29220, 
KP, Pakistan}

\begin{abstract}
Time-resolved Kerr rotation measurements were performed in InGaAs/GaAs quantum wells nearby a doped Mn delta layer. Our magneto-optical results show a typical time evolution of the optically-oriented electron spin in the quantum well. Surprisingly, this is strongly affected by the Mn spins, resulting in an increase of the spin precession frequency in time. This increase is attributed to the variation in the effective magnetic field induced by the dynamical relaxation of the Mn spins. Two processes are observed during electron spin precession: a quasi-instantaneous alignment of the Mn spins with photo-excited holes, followed by a slow alignment of Mn spins with the external transverse magnetic field. The first process leads to an equilibrium state imprinted in the initial precession frequency, which depends on pump power, while the second process promotes a linear frequency increase, with acceleration depending on temperature and external magnetic field. This observation yields new information about exchange process dynamics and on the possibility of constructing spin memories, which can rapidly respond to light while retaining information for a longer period.

\end{abstract}
\begin{document}

\flushbottom
\maketitle

\thispagestyle{empty}

\section*{Introduction}

Magnetic semiconductors can have a huge impact on the market of microelectronic devices, as long as they demonstrate magnetic ordering at room temperature and electron mobility coupled to the magnetic states \cite{Jungwirth2013}. While the improvement of current systems still requires large spin injection, the initial problem of low critical temperature has been solved with the demonstration of $T_{\text{C}} = 250\text{ K}$ in GaAs heterostructures by growing Mn delta-doped layers \cite{Nazmul2005}.

Hybrid structures based on (Ga,Mn)As, combining ferromagnetic (FM) properties with the well-known technology of III-V semiconductor (SC) devices \cite{OhnoScience1998}, are being studied in a bid to increase the number of spin-polarized carriers in semiconductors \cite{HauryPRL1997}. The implementation of a Mn layer in the semiconductor leads to a strong exchange interaction between charge carriers and magnetic atoms at the SC-FM interface, forming a coupled spin system \cite{KorenevIOP2008}. This is particularly appealing since it can be used in applications involving light control of the FM order. This system becomes even richer when the FM layer is placed close to a quantum well, allowing relatively high Curie temperature \cite{Nazmul2005} and hole mobility \cite{Kudrin2018, Dorozhkin2010}. In addition, these heterostructures exhibit strong interaction between carrier spins within the quantum well and the Mn atoms, that allows data storage, as demonstrated in InGaAs/GaAs quantum wells (QWs) adjacent to a Mn delta-doped layer \cite{Balanta2016}. 

The spin interactions in (Ga,Mn)As semiconductors are responsible for the ferromagnetism in these heterostructures and are usually explained by the Zener model \cite{DietlScience2000, Jungwirth2006}. However, the strength of the \textit{sp-d} interaction of the Zener model exponentially decays with distance, while previous studies on InGaAs/GaAs QWs close to Mn delta-layers have demonstrated an influence of the Mn spins on optically-oriented electron spins within QW \cite{Korenev2012, Akimov2014, Zaitsev2016, Balanta2016}, which is not strongly dependent on the distance between the Mn layer and the QW, raising uncertainties about the nature of this interaction.

In this study, carrier spin dynamics measurements were performed by the magneto-optical Kerr technique in InGaAs/GaAs heterostructures nearby doped with a Mn delta layer. We demonstrate the reciprocal possibility of controlling Mn spins by the excited carriers in the QW. Additionally, we observed unexpected acceleration of the spin precession frequency for the electron spins. This phenomenon was attributed to variation in the effective magnetic field produced by the alignment of Mn spins after optical excitation and successive relaxation, which is independent of electron relaxation. This suggests that holes may be responsible for this spin interaction through an effective exchange interaction, as previously observed in other hybrid structures \cite{Korenev2015}. We also show that both the initial alignment of Mn spins with photo-excited holes and the successive alignment with an external magnetic field, can be controlled by experimental parameters, such as pump power, temperature and magnetic field. Furthermore, the Mn spins in our sample respond extremely fast to optical excitation and retain their alignment for longer than the electrons spin coherence,  a condition for spin memory development.

\section{Experimental Measurements}

\begin{figure*}[ht]
  \centering
  \includegraphics[width=170mm]{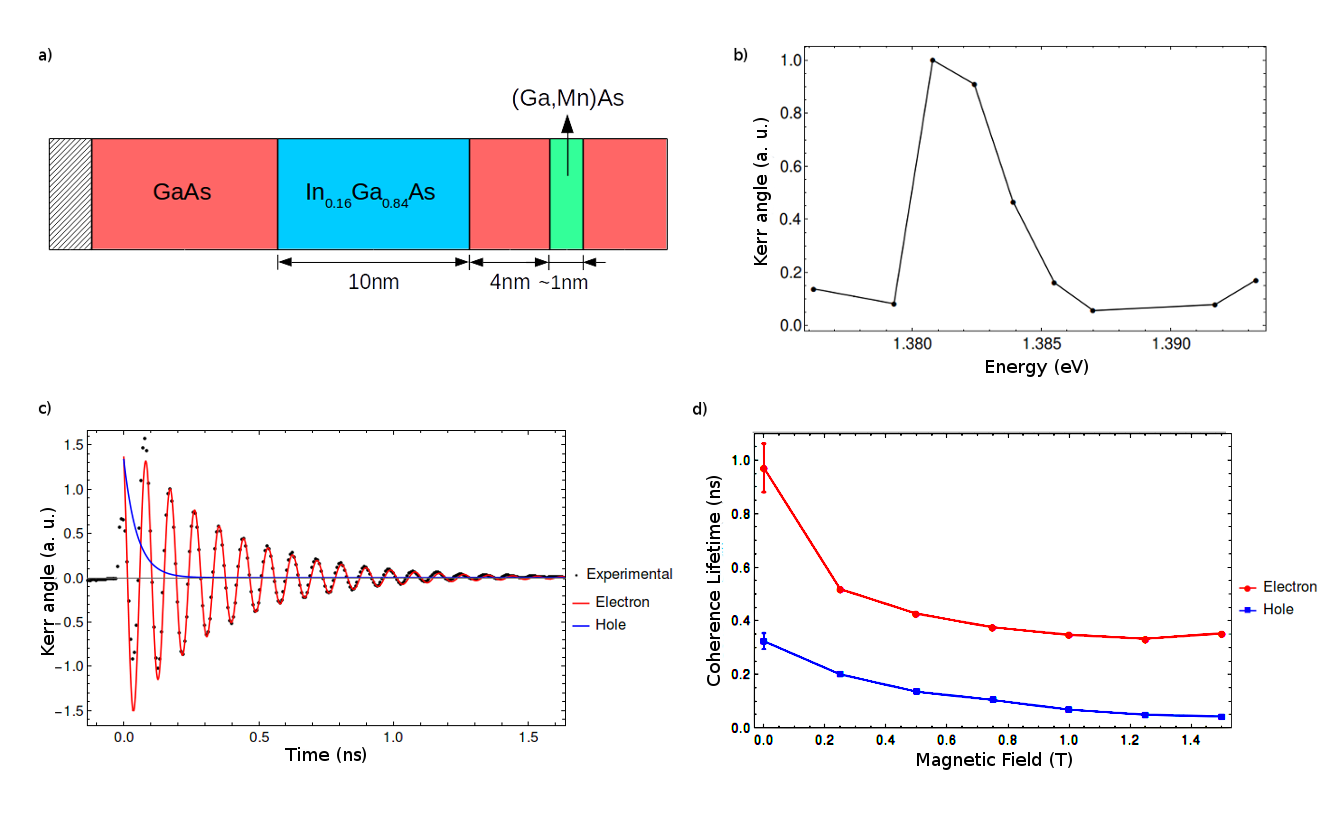}
  \caption{a) GaAs-based heterostructure with a 10 nm InGaAs QW, 4 nm spacer layer and Mn delta-doped layer.
  b) The amplitude of the Kerr rotation signal in arbitrary units for different pump/probe energies with no applied external magnetic field.
  c) Kerr rotation signal (black dots) obtained at $T=5\text{ K}$ with external magnetic field $B_{\text{ext}} = 1.25\text{ T}$. Fits (solid lines) using equation (\ref{Kerr_1}) display a short exponential decay for holes (blue) and a combined oscillation times long decay for electrons (red line).
  d) Coherence time for electron and hole spins at different magnetic fields.}
  \label{MAINFIG}
\end{figure*}

The investigated sample was a $10\text{ nm}$ wide In$_{0.16}$Ga$_{0.84}$As/GaAs QW separated from a Mn delta-doped layer by a $4 \text{ nm}$ spacer (Figure \ref{MAINFIG}a). The Mn deposition forms a thin Mn$_{x}$Ga$_{1-x}$As layer ($\sim1 \text{ nm}$), where the Mn acts as an acceptor impurity, creating a two-dimensional hole gas. The Mn concentration was estimated by the deposition time as $0.3$ monolayers, which may be equivalent to approximately $1.9\times10^{14} \text{ Mn}/\text{cm}^2$ and the Mn incorporation in the GaAs was $x\approx 2-6 \%$. The hole concentration in the QW was obtained by Shubnikov-de-Hass measurements and was approximately $5\times 10^{11} \text{ Mn}\text{ cm}^{-2}$. Further details about the sample MOCVD growth can be obtained elsewhere [\onlinecite{Dorokhin2008}]. The Hall effect and magnetoresistance measurements demonstrate ferromagnetic behaviour of Mn spins with Curie temperature $T_{\text{C}} \approx 35\text{ K}$.

Time-resolved Kerr rotation was performed using a tunable mode-locked Ti:Sapphire laser with a pulse duration of $1 \text{ ps}$ and repetition rate of $76 \text{ MHz}$. The time delay ($t$) between pump and probe pulses was adjusted by a mechanical delay line. The pump beam was circularly polarized by a photoelastic modulator and the probe was linearly polarized and modulated by a chopper. The polarization rotation of the reflected probe beam was detected with a balanced bridge using coupled photodiodes synchronized with pump and probe modulations. The sample was immersed in the variable temperature insert of a split-coil superconductor magnet in the Voigt geometry. The pump and probe energy was tuned to $E\approx1.38\text{ eV}$ at the maximum amplitude of the Kerr signal. Figure \ref{MAINFIG}b shows the excitation energy dependence of the Kerr signal. This energy is consistent with the QW transition identified by PLE measurements in reference [\onlinecite{BalantaJAP2014}]. For this energy, we observed an electron spin coherence time in the order of $1 \text{ ns}$, at $T = 5 \text{ K}$, which is longer than previously reported for similar samples 
\cite{Korenev2012}.

The time-dependent Kerr signal shown in Figure \ref{MAINFIG}c essentially has two oscillations and two exponential decays arising from electron and hole decoherence \cite{Korenev2012, Akimov2014, Zaitsev2016}. The following equation was used to fit experimental data:
\begin{equation}
\theta_{K}=A_e e^{-t/T_2^e}\cos(\omega t + \phi) + A_h e^{-t/T_2^h} 
\label{Kerr_1}
\end{equation} 
where $A_{e,h}$ and $T_2^{e,h}$ are the electron/hole signal amplitude and coherence time, $\omega$ and $\phi$ are the precession frequency and oscillation phase for the electron spins and $t$ is the pump-probe time delay.

In Figure \ref{MAINFIG}d, both electron and hole spin coherence times were found to decrease when the external magnetic field increased, an effect attributed to ensemble g-factor inhomogeneities. The hole precession was not taken into account for fitting due to its short coherence time and much slower spin precession frequency (Figure \ref{MAINFIG}c). The hole and electron parameters in both Figure \ref{MAINFIG}c and Figure \ref{MAINFIG}d were obtained from the fit of equation \ref{Kerr_1} and plotted separately.

The large signal/noise ratio allows the Kerr oscillation to be followed for a broad delay time range. Surprisingly, within this range, phase mismatching was observed between the spin precession data and the fit curve above $0.6\text{ ns}$ (clearer by zooming Kerr signal in this time range). This mismatching could not be resolved by the addition of any other precession, indicating that it is not a consequence of beating. It is more likely that the spin precession velocity is constantly increasing, where this could be a consequence of an increasing effective magnetic field acting on the photo-excited electron spins. This point is the main focus of the present study and is discussed in detail below.

\subsection{Optical magnetization induction}

\begin{figure}[ht!]
  \centering
  \includegraphics[width=85mm]{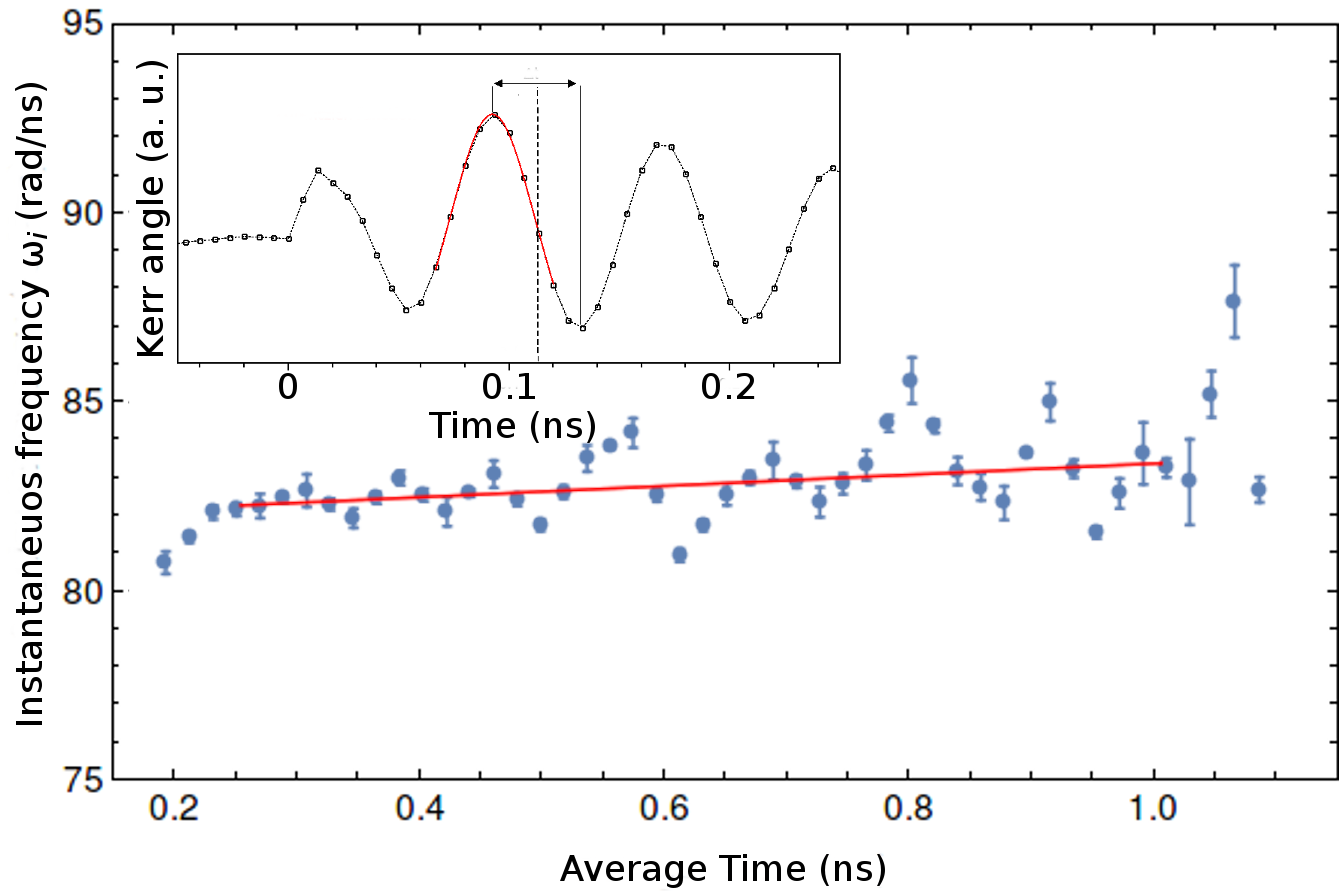}
  \caption{Instantaneous frequency $\omega_{i}$ plotted as function of average time $t_{\text{av}}$ (dots). A linear fit (solid line) shows an average increasing behavior. The position of each peak was obtained by a Lorentzian fit as shown in the inset.}\label{lorentzian_fit}
\end{figure}

In order to better understand the time evolution of the precession frequency and the phase mismatch, the time separation between consecutive maximums and minimums of the Kerr signal $\Delta t_i = t_{\text{max}_i} - t_{\text{min}_i}$ were plotted as a function of the average time $t_{\text{av}} = (t_{\text{max}_i} + t_{\text{min}_i})/2$, where the position of maximums and minimums were obtained from Lorentzian fits (see inset of Figure \ref{lorentzian_fit}). The time separation between peaks was converted into an instantaneous frequency $\omega _i = 1/{\Delta t_i}$ and depicted in Figure \ref{lorentzian_fit}, and indeed was not constant, showing a slightly-crescent average linear behavior.

\begin{figure}[ht!]
  \centering
  \includegraphics[width=85mm]{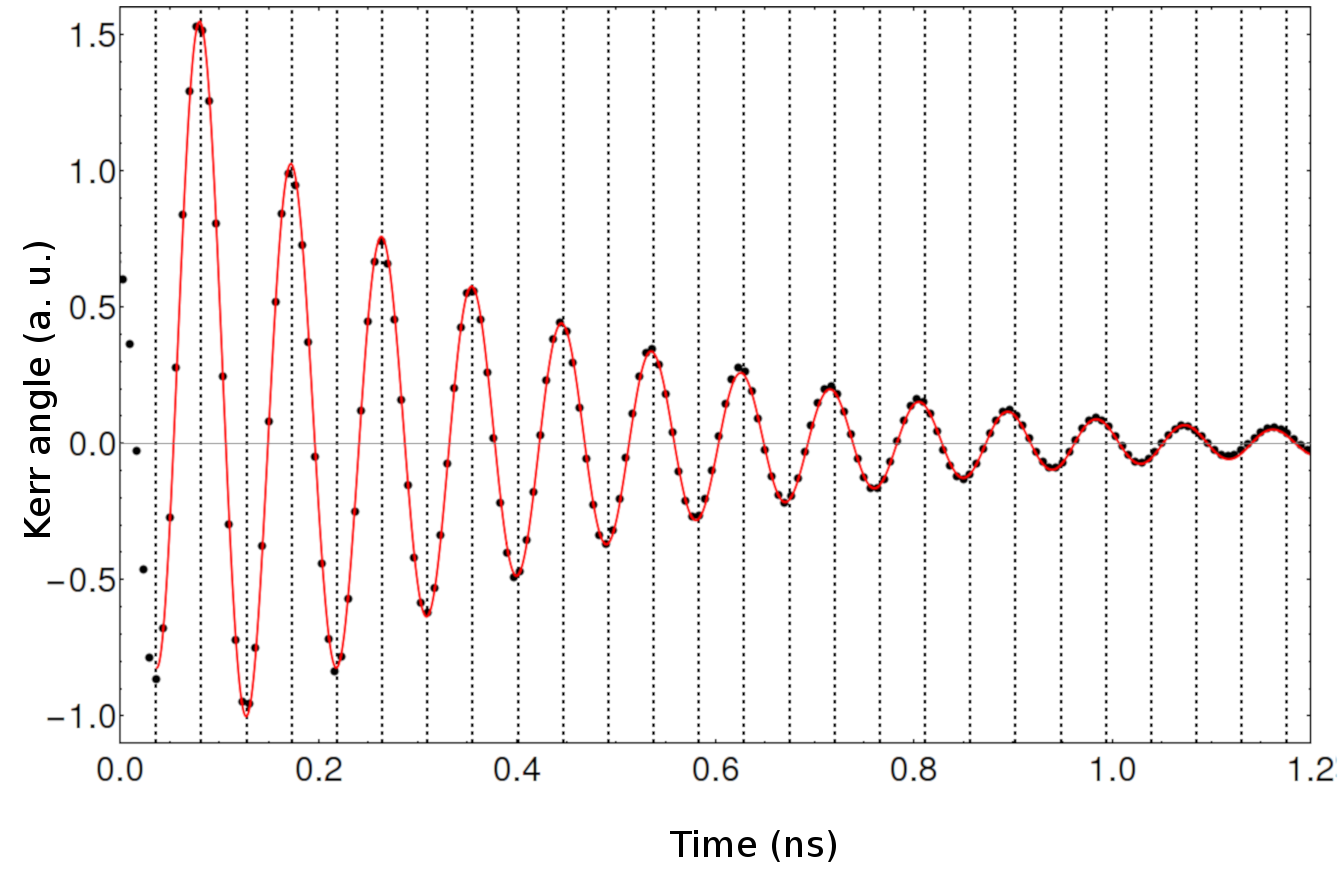}
  \caption{Data fitting of Kerr rotation measurements using equation (\ref{Kerr_2}) at $T=5\text{ K}$ with external magnetic field $B_{\text{ext}} = 1.25\text{ T}$. Vertical grids (equally spaced) show the precession frequency acceleration.}
  \label{new_fit}
\end{figure}

However, since the method described above is based on distance between points, it involves a correlation between consecutive points which may induce an oscillatory behavior. Also, the number of points is limited to the number of oscillations, limiting the precision of the present analysis. Thus, this provides only a rough indicator of average behaviour and, based on the unusual time evolution of the precession frequency, Eq. \ref{Kerr_1} was modified by including a linear time dependent frequency, leading to a new equation:

\begin{equation}
\theta_{K}=A_e e^{-t/T_2^e}\cos(\alpha t^2+\omega_{0}t+\phi)+A_h e^{-t/T_2^h} 
\label{Kerr_2}
\end{equation}
where $\omega_{0}$ is the initial precession frequency, and $\alpha$ is the acceleration factor. From this new equation we can retrieve $\omega_{0}$ and $\alpha$ from original data without the correlation between consecutive points from the previously discussed method. The linear frequency acceleration factor is sufficient to correct the phase mismatching, as evident in Figure \ref{new_fit}, where equally spaced vertical grids were included to highlight frequency acceleration.

The time-dependent behavior of the precession frequency can be assigned to a physical phenomenon similar to that of magnetization induction in the material, as will be consistently demonstrated in the ensuing subsections. There is a strong dependence of $\omega_0 $ and $\alpha$ on magnetic field, temperature and pump power, related to an alignment of the Mn spins induced by the optically-oriented hole spins in the QW, which occurs when the pump pulse creates electrons and holes in the QW. The Mn spins subsequently start to precess and relax, slowly aligning along the external magnetic field. As a consequence, the electrons in the QW are subject to an increase in the transverse component of the effective magnetic field (the constant external magnetic field plus Mn exchange field in the direction of the external field) and start to precess faster. As outlined above, we assume the initial alignment of the Mn spins with the holes is a fast process compared to the subsequent alignment of the Mn spins with the external magnetic field, since no frequency reduction can be observed for short time delays in Figure \ref{lorentzian_fit}. Furthermore, the effective magnetic field on the photo-excited electrons seems to increase linearly  even after the electron spins lose their coherence. This observation is in good concordance with reference [\onlinecite{Balanta2016}], where the effect of a circularly polarized pump on Mn spins is faster than the spectroscopy resolution and remains after exciton recombination.

The first indication that electrons do not significantly affect Mn alignment is that the linear increase of the effective magnetic field is not influenced by the exponential decay of the spin coherence. The assumption that the exchange interaction of the electrons on the Mn spins is not of sufficient strength to affect Mn alignment would explain the inconsistencies observed in previous reports [\onlinecite{Korenev2012, Akimov2014, Zaitsev2016}].

The explanation given above assumes that the optically-oriented hole spins interact strongly with Mn spins causing a quasi-instantaneous alignment. Moreover, the behavior of interaction strength and direction of the initial Mn spin alignment are consistent with a \textit{sp-d} interaction. The reason for the much faster initial alignment is that, for a sufficiently strong pump, the effective field of the coherent holes on the Mn generates a minimum potential, deep enough to flip all Mn spins at the same time. This effective field is much stronger than the effective field generated by the Mn on the photoexcited electrons, as will be demonstrated, since Mn spins do not interact directly with electrons.

Figure \ref{lorentzian_fit} depicts the plot of the experimental and theoretical (Eq. \ref{Kerr_2}) Kerr signals for $B_{\text{ext}} = 1.25 \text{ T}$. The Eq. \ref{Kerr_2} fits  the experimental data relatively well. From the fitting, we extracted the initial frequency $\omega_{0}$ as a function of the pump power, magnetic field and sample temperature. This frequency contains the information on the initial state and interaction forces acting on it.

\subsection{Initial Mn spins orientation}

\begin{figure}[ht]
  \centering
  \includegraphics[width=80mm]{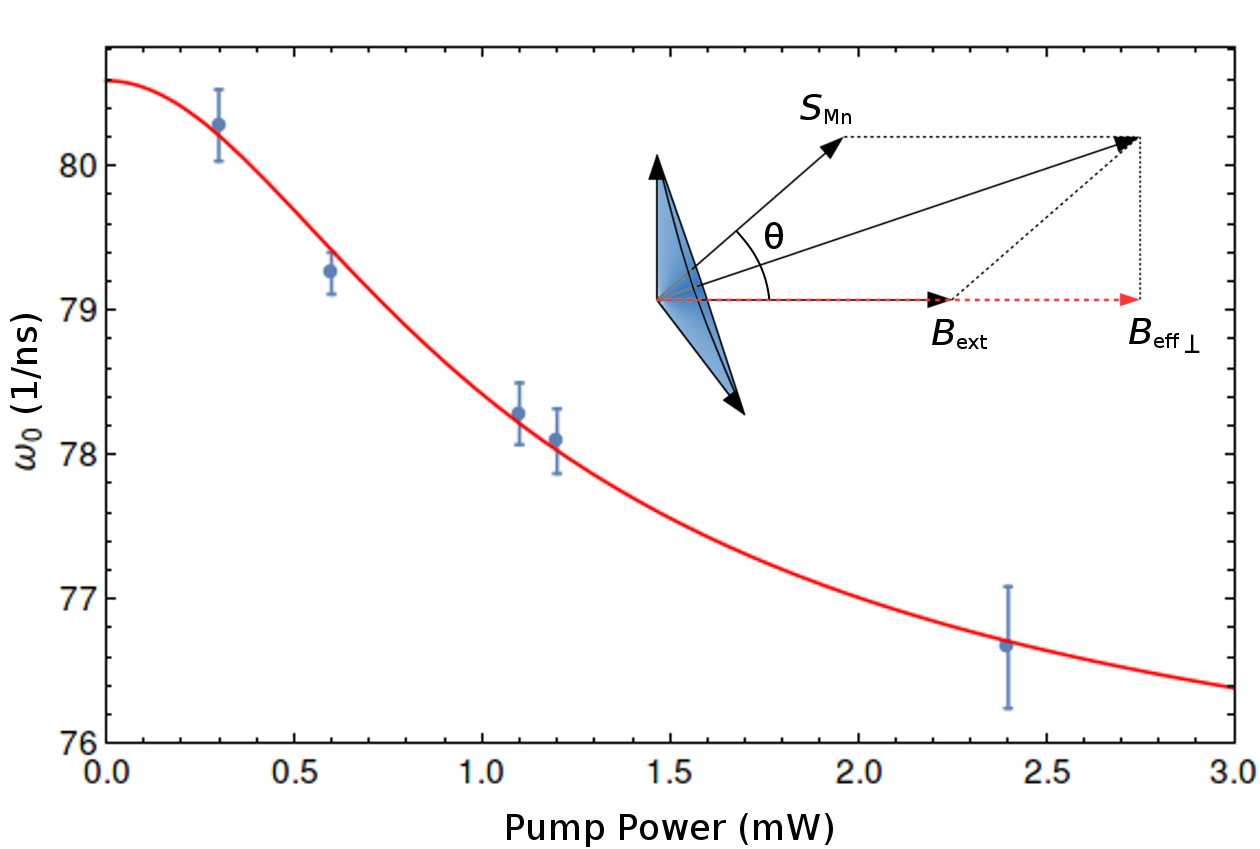}
  \caption{Initial frequency $\omega_0$, extracted from fits of Kerr measurements at $T=5\text{ K}$ and $B=1.5\text{ T}$, as function of pump power (dots), fitted by equation \ref{omega0} (solid line). Inset shows the magnetic field direction scheme and the photoexcited electron spin precession. Since only the spin precession parallel to the pump beam can be measured, only the transverse component of the effective field is relevant.}
  \label{wpp}
\end{figure}

The model, in terms of the spin energy of the system, can be expressed as the sum of two dominant terms: (i) the interaction between Mn spins and the external magnetic field $g_s \mu_{B} (\vec{S}_{\text{Mn}} \cdot 
\vec{B}_{\text{ext}}) = g_s \mu_B S_{\text{Mn}} B_{\text{ext}} \cos \theta$; 
(ii) the exchange interaction between Mn spins and the photo-excited hole spins $\gamma (\vec{S}_{\text{Mn}} \cdot \vec{S}_h) = \gamma S_{\text{Mn}} S_h \sin( \theta)$, where $g_s$ is the Mn \textit{g}-factor, $\mu_B$ is the Bohr magneton, $S_{\text{Mn}}$ and $S_h$ represent the sum of all Mn and hole spins, respectively, while $B_{\text{ext}}$ is the magnitude of the external magnetic field, $\theta$ is the angle between Mn spin and the external field direction (see inset of Figure \ref{wpp}), and $\gamma$ is the strength of the holes-Mn exchange spin interaction. Calculating the equilibrium angle for which the energy is lowest gives: 
\begin{equation}
 \theta_{min} = \text{arctg} \left( \frac{\gamma}{g_s \mu_B} \frac{S_h}{B_{\text{ext}}} \right),
 \label{theta}
\end{equation}
where we can assume that the sum of hole spins $S_h$ is directly proportional to the number of holes and, thus, to the pump power: $S_h=aP$.

Since the effective field of the exchange interaction of Mn spins on the photo-excited electrons is $\frac{\gamma}{g_e \mu_B}S_{\text{Mn}}$, where $g_e$ is the photoexcited electron $g$-factor, and considering only the component of the effective field along the external magnetic field direction, which is the component responsible for the electron precession, gives:

\begin{equation}
 B_{\text{eff}\bot}=B_{\text{ext}}+\frac{\gamma}{g_e \mu_B}S_{\text{Mn}}\cos\theta.
 \label{Bperp}
\end{equation}

Finally, the initial photo-excited electron spin precession frequency can be derived from \ref{theta} and \ref{Bperp}:

\begin{equation}
 \omega_0 = \frac{g_e \mu_B B_{\text{ext}}}{\hbar}+\frac{\gamma S_{\text{Mn}}}{\hbar}\cos \left[ \text{arctg} \left( \frac{\gamma}{g_s \mu_B} \frac{aP}{B_{\text{ext}}} \right) \right].
 \label{omega0}
\end{equation}

Figure \ref{wpp}, depicts the pump-power dependence of the initial frequency $\omega_0$ for $B_{\text{ext}}=1.5\text{ T}$. The Eq. \ref{omega0} precisely fits the data of $\omega_0$, allowing independent extraction of the electron $g$-factor $g_e=0.57$, the effective field of the Mn spins on the photo-excited electron spins $\gamma S_{\text{Mn}} / (g_e \mu_B) = 110 \text{ mT}$ and the effective field of the photo-excited hole spins on the Mn spins as a function of pump power  $\gamma a / (g_s \mu_B) = 1.93 \text{ T} / \text{mW}$.

\subsection{Spin dynamics}

\begin{figure}[ht]
  \centering
  \includegraphics[width=80mm]{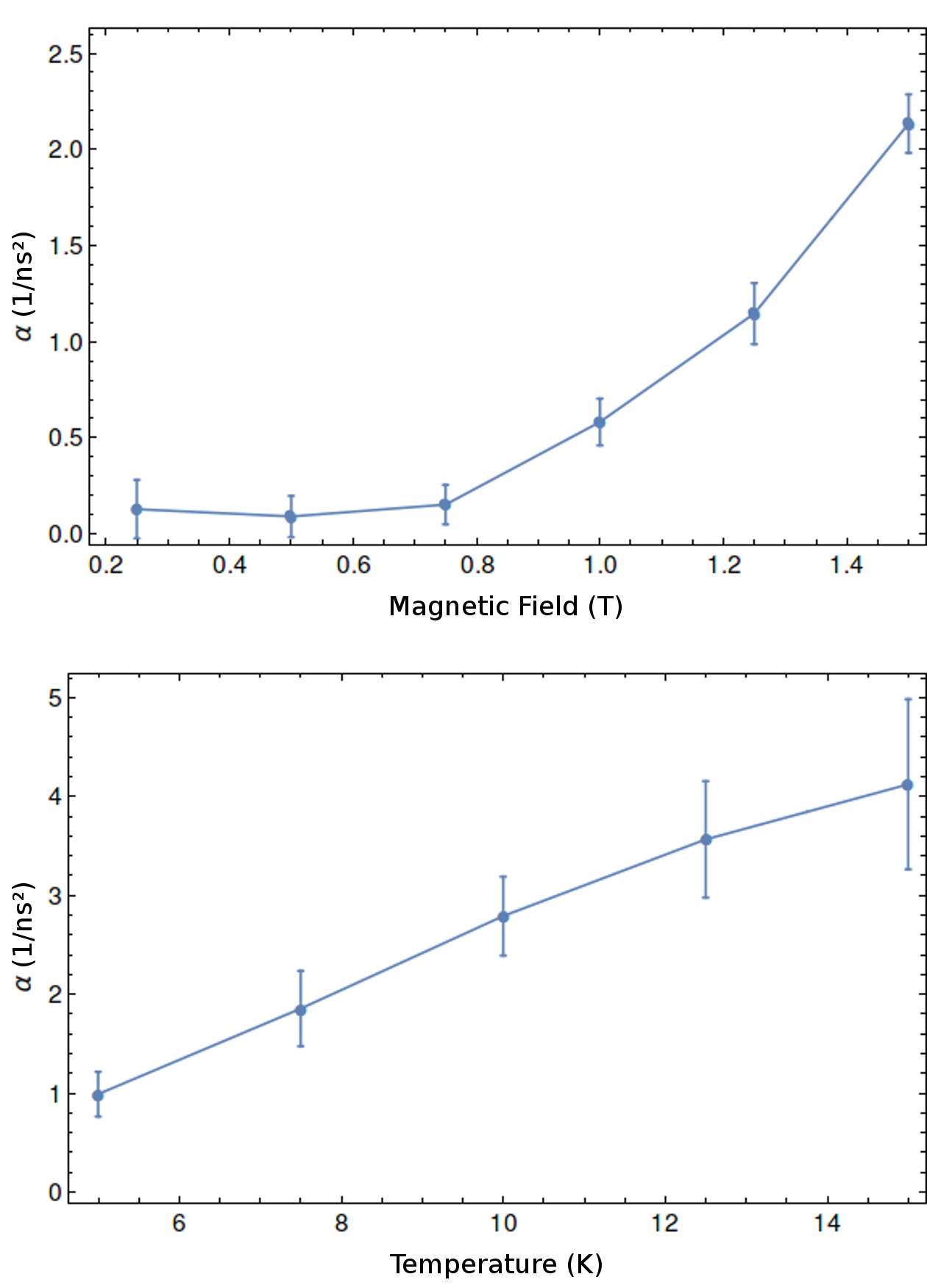}
  \caption{a) Frequency acceleration factor as function of external magnetic field at $T=5\text{ K}$ with pump power $P=2.4\text{ mW}$. b) Frequency acceleration factor as function of temperature at $B=1.5\text{ T}$ with pump power $P=2.4\text{ mW}$.}
  \label{alpha}
\end{figure}

While $\omega_0$ yields information about the equilibrium condition when $t \rightarrow 0$, $\alpha$ can elucidate the dynamics of the system reorientation. Since $\alpha$ can be extracted by fitting Eq. \ref{Kerr_2} with the  experimental data, its values were obtained for different applied magnetic fields, temperatures, and pump power. Figure \ref{alpha}, depicts the plot of $\alpha$ versus $B_{\text{ext}}$ and temperature. This shows strong dependence on the external magnetic field (Figure \ref{alpha}a), as expected, since the magnetic field acts as a restoring force. However, the observed increase of $\alpha$ with the temperature (Figure \ref{alpha}b) is apparently was unexpected. Nevertheless, this can be understood by making an analogy with the spin-lattice relaxation process observed in nuclear magnetic resonance \cite{Bloch1946, Bloembergen1948}. When electron spins precess around the effective magnetic field they lose coherence due to a spin-spin relaxation mechanism, characterized by the spin-spin relaxation time, $T_2$. At the same time, the Mn spins relax and start to align in the direction of the static magnetic field through a spin-lattice relaxation mechanism, characterized by the spin-lattice relaxation time, $T_1$. Due to the Mn spin relaxation, the transverse component of the effective magnetic field increases and tilts in the direction of the external field, increasing the electron precession frequency. Since $\alpha$ is the acceleration associated with the spin-lattice relaxation time, $T_1$ is shorter for larger $\alpha$ values. In this dynamic system, Mn ferromagnetism forces all Mn spins to oscillate together, preventing possible inhomogeneities in the doped layer. When the temperature is increased, new orientation states become accessible to the Mn spins which may lose their coherence much faster, tending to realign with the external field. Consequently, higher temperatures induce larger acceleration and faster decoherence rates, as observed in the study. This phenomenon occurs up to Curie temperature.

\section{Conclusions}

In the current time-dependent magneto-optical studies of the InGaAs/GaAs QWs doped with a Mn delta-layer, there was an increase in time of the electron spin precession frequency in the QW. These results suggest that the Mn spins are affected by the pump beam, due to an interaction with photoexcited holes, and also further relax, aligning with the external magnetic field. After holes lose coherence, the Mn spins interact with the photoexcited electrons as an effective field, combining with the external field and making electrons precess faster. These two process mechanisms (Mn spin orientation with pump and further relaxation) occur at different velocities.  This can explain inconsistencies observed in previous studies of the same sample between Kerr rotation and photoluminescence results \cite{Korenev2012, Akimov2014, Zaitsev2016}, where the Mn field strength, obtained by Kerr rotation frequency when the magnetic field tends to zero, is not sufficiently strong to explain the magnetization of the sample, measured by photoluminescence.

Despite the extremely fast initial alignment of the Mn spins, the forces involved in this process can be studied and separated from the Mn effective field, assuming that the system is initially at equilibrium. Our results show that the field responsible for the initial alignment depends on the pump power and, at $\text{ T} = 5\text{ K}$, for $1\text{ mW}$, it is almost 20 times stronger than the effective Mn field. This disparity in strength, together with the time difference between the two alignment processes, also indicates that the initial alignment may not be related to a direct interaction between Mn spins and electrons in the QW, since the initial alignment remains even after the electron loses coherence. On the other hand, photoexcited holes lose coherence extremely fast, where this may be related to the initial alignment via an effective \textit{sp-d} interaction.

The different time scales for the two dynamical processes are very attractive and make these structures a promising system for technological applications. The sample response for external control with light can be extremely fast, while the spin memory may be long. Apart from the particular properties of our sample, the method proposed here to explain and analyze mismatches in the fitting model can be extended to many other systems and yield relevant information about the dynamic of different spin processes.

\bibliography{sample}

\begin{thebibliography}{10}
\urlstyle{rm}
\expandafter\ifx\csname url\endcsname\relax
  \def\url#1{\texttt{#1}}\fi
\expandafter\ifx\csname urlprefix\endcsname\relax\def\urlprefix{URL }\fi
\expandafter\ifx\csname doiprefix\endcsname\relax\def\doiprefix{DOI: }\fi
\providecommand{\bibinfo}[2]{#2}
\providecommand{\eprint}[2][]{\url{#2}}

\bibitem{Jungwirth2013}
\bibinfo{author}{Jungwirth, T.} \emph{et~al.}
\newblock \bibinfo{journal}{\bibinfo{title}{Dc-transport properties of
  ferromagnetic {(Ga,Mn)As} semiconductors}}.
\newblock {\emph{\JournalTitle{Applied Physics Letters}}}
  \textbf{\bibinfo{volume}{83}}, \bibinfo{pages}{320--322},
  \doiprefix\url{10.1063/1.1590433} (\bibinfo{year}{2003}).

\bibitem{Nazmul2005}
\bibinfo{author}{Nazmul, A.~M.}, \bibinfo{author}{Amemiya, T.},
  \bibinfo{author}{Shuto, Y.}, \bibinfo{author}{Sugahara, S.} \&
  \bibinfo{author}{Tanaka, M.}
\newblock \bibinfo{journal}{\bibinfo{title}{High temperature ferromagnetism in
  {GaAs}-based heterostructures with{ Mn} $\ensuremath{\delta}$ doping}}.
\newblock {\emph{\JournalTitle{Phys. Rev. Lett.}}}
  \textbf{\bibinfo{volume}{95}}, \bibinfo{pages}{017201},
  \doiprefix\url{10.1103/PhysRevLett.95.017201} (\bibinfo{year}{2005}).

\bibitem{OhnoScience1998}
\bibinfo{author}{Ohno, H.}
\newblock \bibinfo{journal}{\bibinfo{title}{Making nonmagnetic semiconductors
  ferromagnetic}}.
\newblock {\emph{\JournalTitle{Science}}} \textbf{\bibinfo{volume}{281}},
  \bibinfo{pages}{951--956}, \doiprefix\url{10.1126/science.281.5379.951}
  (\bibinfo{year}{1998}).

\bibitem{HauryPRL1997}
\bibinfo{author}{Haury, A.} \emph{et~al.}
\newblock \bibinfo{journal}{\bibinfo{title}{Observation of a ferromagnetic
  transition induced by two-dimensional hole gas in modulation-doped {CdMnTe}
  quantum wells}}.
\newblock {\emph{\JournalTitle{Phys. Rev. Lett.}}}
  \textbf{\bibinfo{volume}{79}}, \bibinfo{pages}{511--514},
  \doiprefix\url{10.1103/PhysRevLett.79.511} (\bibinfo{year}{1997}).

\bibitem{KorenevIOP2008}
\bibinfo{author}{Korenev, V.~L.}
\newblock \bibinfo{journal}{\bibinfo{title}{Optical orientation in
  ferromagnet/semiconductor hybrids}}.
\newblock {\emph{\JournalTitle{Semiconductor Science and Technology}}}
  \textbf{\bibinfo{volume}{23}}, \bibinfo{pages}{114012}
  (\bibinfo{year}{2008}).

\bibitem{Kudrin2018}
\bibinfo{author}{Kudrin, A.~V.} \emph{et~al.}
\newblock \bibinfo{title}{The nature of transport and ferromagnetic properties
  of the {G}a{A}s structures with the {M}n $\delta$-doped layer}
  (\bibinfo{year}{2018}).
\newblock \eprint{arXiv:1804.07650}.

\bibitem{Dorozhkin2010}
\bibinfo{author}{Dorozhkin, S.~I.} \emph{et~al.}
\newblock \bibinfo{journal}{\bibinfo{title}{Beatings of {S}hubnikov-de {H}aas
  oscillations in a two-dimensional hole system in an {I}n{G}a{A}s quantum
  well}}.
\newblock {\emph{\JournalTitle{JETP Letters}}} \textbf{\bibinfo{volume}{91}},
  \bibinfo{pages}{292--296}, \doiprefix\url{10.1134/S002136401006007X}
  (\bibinfo{year}{2010}).

\bibitem{Balanta2016}
\bibinfo{author}{Balanta, M. A.~G.} \emph{et~al.}
\newblock \bibinfo{journal}{\bibinfo{title}{Optically controlled
  spin-polarization memory effect on {Mn} delta-doped heterostructures}}.
\newblock {\emph{\JournalTitle{Scientific Reports}}}
  \textbf{\bibinfo{volume}{6}}, \bibinfo{pages}{24537} (\bibinfo{year}{2016}).

\bibitem{DietlScience2000}
\bibinfo{author}{Dietl, T.}, \bibinfo{author}{Ohno, H.},
  \bibinfo{author}{Matsukura, F.}, \bibinfo{author}{Cibert, J.} \&
  \bibinfo{author}{Ferrand, D.}
\newblock \bibinfo{journal}{\bibinfo{title}{Zener model description of
  ferromagnetism in zinc-blende magnetic semiconductors}}.
\newblock {\emph{\JournalTitle{Science}}} \textbf{\bibinfo{volume}{287}},
  \bibinfo{pages}{1019--1022}, \doiprefix\url{10.1126/science.287.5455.1019}
  (\bibinfo{year}{2000}).
\newblock
  \eprint{http://science.sciencemag.org/content/287/5455/1019.full.pdf}.

\bibitem{Jungwirth2006}
\bibinfo{author}{Jungwirth, T.}, \bibinfo{author}{Sinova, J.},
  \bibinfo{author}{Ma\ifmmode~\check{s}\else \v{s}\fi{}ek, J.},
  \bibinfo{author}{Ku\ifmmode~\check{c}\else \v{c}\fi{}era, J.} \&
  \bibinfo{author}{MacDonald, A.~H.}
\newblock \bibinfo{journal}{\bibinfo{title}{Theory of ferromagnetic {(III,Mn)V}
  semiconductors}}.
\newblock {\emph{\JournalTitle{Rev. Mod. Phys.}}}
  \textbf{\bibinfo{volume}{78}}, \bibinfo{pages}{809--864},
  \doiprefix\url{10.1103/RevModPhys.78.809} (\bibinfo{year}{2006}).

\bibitem{Korenev2012}
\bibinfo{author}{Korenev, V.} \emph{et~al.}
\newblock \bibinfo{journal}{\bibinfo{title}{Dynamic spin polarization by
  orientation-dependent separation in a ferromagnet--semiconductor hybrid}}.
\newblock {\emph{\JournalTitle{Nature communications}}}
  \textbf{\bibinfo{volume}{3}}, \bibinfo{pages}{959},
  \doiprefix\url{10.1038/ncomms1957} (\bibinfo{year}{2012}).

\bibitem{Akimov2014}
\bibinfo{author}{Akimov, I.~A.} \emph{et~al.}
\newblock \bibinfo{journal}{\bibinfo{title}{Orientation of electron spins in
  hybrid ferromagnet–semiconductor nanostructures}}.
\newblock {\emph{\JournalTitle{physica status solidi (b)}}}
  \textbf{\bibinfo{volume}{251}}, \bibinfo{pages}{1663--1672},
  \doiprefix\url{10.1002/pssb.201350236} (\bibinfo{year}{2014}).

\bibitem{Zaitsev2016}
\bibinfo{author}{Zaitsev, S.~V.} \emph{et~al.}
\newblock \bibinfo{journal}{\bibinfo{title}{Coherent spin dynamics of carriers
  in ferromagnetic semiconductor heterostructures with an {Mn} $\delta$
  layer}}.
\newblock {\emph{\JournalTitle{Journal of Experimental and Theoretical
  Physics}}} \textbf{\bibinfo{volume}{123}}, \bibinfo{pages}{420--428},
  \doiprefix\url{10.1134/S106377611607013X} (\bibinfo{year}{2016}).

\bibitem{Korenev2015}
\bibinfo{author}{{Korenev V. L.}} \emph{et~al.}
\newblock \bibinfo{journal}{\bibinfo{title}{{Long-range p--d exchange
  interaction in a ferromagnet--semiconductor hybrid structure}}}.
\newblock {\emph{\JournalTitle{Nature Physics}}} \textbf{\bibinfo{volume}{12}},
  \bibinfo{pages}{85--91}, \doiprefix\url{http://dx.doi.org/10.1038/nphys3497
  10.1038/nphys3497} (\bibinfo{year}{2016}).

\bibitem{Dorokhin2008}
\bibinfo{author}{Dorokhin, M.~V.} \emph{et~al.}
\newblock \bibinfo{journal}{\bibinfo{title}{Emission properties of
  {InGaAs/GaAs} heterostructures with delta {Mn}-doped barrier}}.
\newblock {\emph{\JournalTitle{Journal of Physics D: Applied Physics}}}
  \textbf{\bibinfo{volume}{41}}, \bibinfo{pages}{245110}
  (\bibinfo{year}{2008}).

\bibitem{BalantaJAP2014}
\bibinfo{author}{Balanta, M. A.~G.} \emph{et~al.}
\newblock \bibinfo{journal}{\bibinfo{title}{Effects of a nearby {Mn} delta
  layer on the optical properties of an {InGaAs/GaAs} quantum well}}.
\newblock {\emph{\JournalTitle{Journal of Applied Physics}}}
  \textbf{\bibinfo{volume}{116}}, \bibinfo{pages}{203501},
  \doiprefix\url{10.1063/1.4902857} (\bibinfo{year}{2014}).

\bibitem{Bloch1946}
\bibinfo{author}{Bloch, F.}
\newblock \bibinfo{journal}{\bibinfo{title}{Nuclear induction}}.
\newblock {\emph{\JournalTitle{Phys. Rev.}}} \textbf{\bibinfo{volume}{70}},
  \bibinfo{pages}{460--474}, \doiprefix\url{10.1103/PhysRev.70.460}
  (\bibinfo{year}{1946}).

\bibitem{Bloembergen1948}
\bibinfo{author}{Bloembergen, N.}, \bibinfo{author}{Purcell, E.~M.} \&
  \bibinfo{author}{Pound, R.~V.}
\newblock \bibinfo{journal}{\bibinfo{title}{Relaxation effects in nuclear
  magnetic resonance absorption}}.
\newblock {\emph{\JournalTitle{Phys. Rev.}}} \textbf{\bibinfo{volume}{73}},
  \bibinfo{pages}{679--712}, \doiprefix\url{10.1103/PhysRev.73.679}
  (\bibinfo{year}{1948}).

\end{thebibliography}

\section*{Acknowledgements}

We acknowledge financial support from Grants No 2015/16191-5 and 2016/16365-6 of the S\~{a}o Paulo Research Foundation (FAPESP), Grant No 305769/2015-4 of the Coordination for the Improvement of Higher Education Personnel (CAPES) and Grant No. 8.1751.2017/PCh of the Ministry of Education and Science of Russian Federation

\section*{Author contributions statement}

F.C.D.M., F.G.G.H., M.G.A.B. and F.I. conceived the experiments; Y.A.D., M.V.D., O.V.V and B.N.Z. fabricated the sample, carried out sample preliminary characterization and participated in the discussion; F.C.D.M. and S.U. conducted the experiments; F.C.D.M. analyzed the results and proposed the theoretical model; F.C.D.M., F.G.G.H., M.G.A.B. and F.I. discussed the results, F.M wrote the manuscript and all authors reviewed it.

\section*{Additional Information}
\textbf{Competing interests:} the authors declare no competing interests.

\end{document}